\documentclass{jpp}
\usepackage{graphicx}

\usepackage[utf8]{inputenc}
\usepackage[T1]{fontenc}
\usepackage{amsmath}
\usepackage{url}

\usepackage{natbib}
\newcommand{\be}{\begin{displaymath}}
\newcommand{\bn}{\begin{equation}}

\newcommand{\en}{\end{equation}}
\newcommand{\ee}{\end{displaymath}}
\newcommand{\lang}{\left\langle}
\newcommand{\rang}{\right\rangle}

\newcommand{\simlt}{\:{\raisebox{-1.5mm}{$\stackrel
{\textstyle{<}}{\sim}$}}\:}

\shorttitle{Shafranov shift in stellarators}
\shortauthor{P. Helander and N. Nikulsin}

\title{On the Shafranov shift in stellarators}

\author{Per Helander
  \corresp{\email{per.helander@ipp.mpg.de}} \and Nikita Nikulsin}

\affiliation{Max Planck Institute for Plasma Physics, Greifswald Germany}

\begin{document}

\maketitle

\begin{abstract}
As first shown by Shafranov, toroidal plasmas in magnetohydrodynamic equilibrium tend to expand in major radius when the pressure is increased. Here, an average measure of the resulting Shafranov shift is introduced, and its properties are discussed for various classes of stellarator configurations. It is shown to be particularly small in quasi-helical and quasi-isodynamic stellarators with a large number of field periods, which are thus relatively robust to variations in the plasma pressure. 
\end{abstract}

\section{Introduction}

Vitaly Dmitrievich Shafranov noted that toroidal plasmas in magnetohydrodynamic (MHD) equilibrium have the tendency to move into regions of weaker magnetic field when the pressure is increased. He calculated the resulting shift of the magnetic surfaces in a large-aspect-ratio tokamak with circular cross section \citep{Shafranov-1958,Shafranov-1966} -- a calculation that is reproduced in most textbooks on fusion plasma physics. He also calculated the corresponding shift in a stellarator with almost circular cross section  \citep{Shafranov-1966,Solovev-1970}, but most stellarators are strongly shaped and this circumstance can make the Shafranov shift quite different.

The Shafranov shift is of considerable practical importance in stellarators, because it sets a limit on the achievable pressure -- an MHD equilibrium limit rather than a stability limit. At high enough beta (thermal pressure divided by magnetic energy density), the Shafranov shift is comparable to the minor radius of the plasma, causing the latter to be pushed against the outer boundary and a magnetic separatrix to form that limits the confinement volume. Alternatively, a region of chaotic field lines can appear in the periphery of the plasma, also reducing confinement volume \citep{Drevlak_2005,Loizu-2017}. In addition to causing an MHD equilibrium limit, the Shafranov shift can greatly enhance neoclassical transport and fast-particle losses. 

\cite{Spitzer} estimated the pressure limit in a classical stellarator with circular cross section to scale as
    \bn \beta \sim \epsilon \iota^2, 
    \label{classical}
    \en
where $\epsilon$ denotes the inverse aspect ratio and $\iota$ the rotational transform. Many other special cases have been calculated more carefully over the years \citep{Pustovitov}, and it would be useful to understand the Shafranov shift in an arbitrarily shaped stellarator. However, there is no general analytical formula for it, and it is difficult to see how there could be one. With increasing pressure, each point on the magnetic axis is displaced in some direction and by some distance that both vary along the magnetic axis. Moreover, these quantities depend on the entire pressure profile and the boundary condition at the edge of the plasma. In the present article, instead of considering this level of detail, we introduce an {\em average} measure of the Shafranov shift and discuss its properties in various classes of magnetic configurations, particularly quasisymmetric and quasi-isodynamic ones. In order to make analytical progress, it is necessary to adopt the approximation of large aspect ratio, but the shape of the magnetic axis and the plasma cross section are taken to be arbitrary. As a result, we are able to derive scalings and upper bounds on the Shafranov shift of more general nature than has earlier been possible.\footnote{To some extent, general expressions for the Shafranov shift have also been calculated by expanding the MHD equilibrium equations around the magnetic axis \citep{Landreman_2021,Rodriguez_2023}, which mirrors our large-aspect-ratio expansion, but the latter is somewhat more general as it allows the plasma cross section to be arbitrarily shaped whereas near-axis equilibria have approximately elliptical cross section.}

\section{Magnetic field and plasma current}

\subsection{General expressions}

In this section, we recall general expressions for the magnetic field and plasma current in stellarators. Most of the results are well known and can be found in several places of the published literature, see e.g. the review by \cite{helander2014}. The plasma is assumed to be in a scalar-pressure MHD equilibrium without flow, in which the magnetic field lines trace out simply nested flux surfaces, 
	$$ {\bf J} \times {\bf B} = \nabla p, $$
where $p$ denotes the pressure, ${\bf B}$ the magnetic field and ${\bf J}$ the current density. Boozer coordinates $(\psi,\theta,\varphi)$ can then be introduced and the magnetic field written as \citep{Boozer-1981}
	\bn
	 {\bf B} = \nabla \psi \times \nabla \theta + \nabla \varphi \times \nabla \chi = G(\psi) \nabla \varphi + I(\psi) \nabla \theta + K(\psi,\theta,\varphi) \nabla \psi, 
	\label{B in magnetic coordinates}
	\en
where $\psi$ and $\chi$ measure toroidal and poloidal flux, respectively. The quantity $\psi$ is chosen to vanish on the magnetic axis, while $\chi$ is taken to vanish at the plasma edge and thus to attain its maximum\footnote{or minimum, depending on whether $\iota = d\chi/d\psi$ is positive or negative} on the magnetic axis. The function $I(\psi)$ is proportional to the total toroidal current enclosed by the flux surface $\psi$. In most stellarators $I(\psi)$ is much smaller than $G(\psi)$, since the current in the toroidal-field coils is much larger than the toroidal plasma current. The derivative $G'(\psi)$ is proportional to the pressure gradient, making $G(\psi)$ almost constant in a low-beta plasma. The Jacobian is
	$$ \frac{1}{\sqrt{g}} = (\nabla \psi \times \nabla \theta) \cdot \nabla \varphi = \frac{B^2}{G + \iota I}, $$
where $\iota(\psi) = d \chi / d \psi$ denotes the rotational transform. We shall also use another set of coordinates $(\psi,\alpha,l)$, where $\alpha = \theta - \iota(\psi) \varphi$ labels the different field lines on each flux surface, and $l$ denotes the arc length along each field line. 

The flux-surface average of a function $f(\psi,\theta,\varphi)$ is defined as the volume-average of this function over the region between two neighbouring magnetic surfaces, 
	$$ \lang f \rang = \frac{1}{V'(\psi)} \int_0^{2\pi} d\theta \int_0^{2\pi/N_p} f(\psi,\theta,\varphi) \sqrt{g(\psi,\theta,\varphi)} \; d\varphi, $$
where 
	$$ V'(\psi) = \int_0^{2\pi} d\theta \int_0^{2\pi/N_p} \sqrt{g} \; d\varphi $$
denotes the volume enclosed by the surface labelled by $\psi$ in one period of the stellarator, whose shape is assumed to be invariant under toroidal rotation by the angle $2 \pi / N_p$. Here, $N_p$ denotes the number of field periods. 

Since the divergence of the diamagnetic current
	$$ {\bf J}_\perp = \frac{{\bf B} \times \nabla p}{B^2} $$
in general does not vanish, already \cite{Spitzer} noted that there must be a current along $\bf B$ satisfying 
	\bn {\bf B} \cdot \nabla \left(\frac{J_\|}{B} \right) = - ({\bf B} \times \nabla p) \cdot \nabla \left( \frac{1}{ B^{2}} \right). 
	\label{magnetic DE}
	\en
The Pfirsch-Schl{\"u}ter current is defined as the part of this current	that does not contribute to the net toroidal current, i.e. 
	$$ J_{\|}^{\rm PS} = J_{\|} - \frac{\lang J_{\|} B \rang B}{\lang B^2 \rang}. $$
In the absence of an Ohmic current, the bootstrap current, and auxilliary current drive, $J_{\|}^{\rm PS} = J_{\|}$. As Boozer first observed, the Pfirsch-Schl{\"u}ter current can be expressed in terms of the Fourier coefficients of 
	$$ \frac{1}{B^2} = \sum_{m,n} h_{mn}(\psi) e^{i(m \theta - n \varphi)},  $$
as
    \bn \frac{J_\|^{\rm PS}}{B} = -\frac{dp}{d\psi} \sum_{m,n\ne 0} \frac{nI + mG}{m \iota - n} 
    h_{mn} e^{i(m \theta - n \varphi)}, 
	\label{JPS-exact}
	\en
where each coefficient $h_{mn}(\psi)$ vanishes on the resonant surface where $\iota(\psi)=n/m$. Otherwise, a current singularity arises, which has recently been analysed in detail by \cite{Huang}.

\subsection{Quasisymmetric magnetic fields}

Quasisymmetric fields are those for which the field strength only depends on a linear combination of the Boozer angles, $B = B(\psi,\theta - N \varphi)$, where the integer $N$ vanishes in quasi-axisymmetric stellarators and usually coincides with the  number of field periods in quasihelically symmetric ones. For such fields, any function of the form $f(\psi,B)$ satisfies
	$$ {\bf B} \cdot \nabla f = \frac{\iota - N}{\sqrt{g}} 
	\left( \frac{\partial f}{\partial \theta} \right)_{\psi,\varphi}, $$
	$$ ({\bf B} \times \nabla \psi) \cdot \nabla f = \frac{G + NI}{\sqrt{g}}
	\left( \frac{\partial f}{\partial \theta} \right)_{\psi,\varphi}, $$
and (\ref{magnetic DE}) reduces to an ordinary differential equation that can be solved analytically, giving
	\bn	J_{\|}^{\rm PS} = \frac{G + NI}{N-\iota} \frac{dp}{d\psi} \left( \frac{1}{B} - \frac{B}{\lang B^2 \rang} \right). \label{QS JPS}
	\en
Note that, in a tokamak or a quasi-axismmetric stellarator ($N = 0$), the Pfirsch-Schl{\"u}ter current is proportional to $G/\iota$, whereas in a quasihelically symmetric stellarator without net current ($I = 0$), it is instead proportional to $G/(N - \iota)$. It is thus relatively small if the number of field periods is large. 

\subsection{Quasi-isodynamic magnetic fields}

By definition, a quasi-isodynamic magnetic field satisfies the following two conditions \citep{Helander-2009,Nuhrenberg-2010}. 
\begin{itemize}

	\item[(i)] On each flux surface, all contours of constant field strength have the same topology and form poloidally closed curves, rather than toroidally or helically closed ones. 
	
	\item[(ii)] For trapped particles, the distance along the field between consecutive bounce points depends on the field strength at these points and on the flux surface on which they reside, but is the same for all field lines on that surface. In other words, in the coordinates $(\psi,\alpha,l)$ introduced above, the arc length between two points, $(\psi,\alpha,l_0)$ and $(\psi,\alpha,l_1)$, of equal field strength on the same field line depends on $B(\psi,\alpha,l_0) = B(\psi,\alpha,l_1)$ and $\psi$ but not on the Clebsch angle $\alpha$, provided that $B(\psi,\alpha,l) < B(\psi,\alpha,l_0) $ for all $l \in (l_0,l_1)$. In particular, the distance between two consecutive field-strength maxima,
		$$ L(\psi) = \int_{B(\psi,\alpha,l) \le B_{\rm max}(\psi)} dl, $$
depends only on $\psi$.

\end{itemize}

These conditions, which have been elaborated upon by \cite{Parra-2015} and \cite{Helander-2024}, guarantee that magnetically trapped particles, on a time average, drift in the poloidal direction rather than radially, toroidally or helically \citep{Cary-1997}. They also ensure that, for any function $f(\psi,B)$, 
	   \bn \frac{\partial}{\partial \alpha} \int_{B(\psi,\alpha,l)<B} f[\psi,B(\psi,\alpha,l)] dl = 0. 
       \label{CS}
       \en
       
It was observed by \cite{Subbotin-2006} that this property has an important implication for the Pfirsch-Schl{\"u}ter current in the important case that the net toroidal current vanishes, $I(\psi) = 0$, so that ${\bf B} = G \nabla \varphi + K \nabla \psi$. In the coordinate system $(\psi,\alpha, l)$, the equation for the Pfirsch-Schl{\"u}ter current (\ref{magnetic DE}) then becomes
	\bn \frac{\p u}{\p l} = - 2 \; \frac{\p}{\p \alpha} \left( \frac{1}{B} \right), 
    \label{eq for u}
    \en
where $u(\psi,\alpha,l) = J_\| / [p'(\psi) B]$. If we integrate this equation along the field over one period of the stellarator, from one field maximum to the next\footnote{There could, in principle, be several field maxima \citep{Parra-2015} in each period, but we integrate between the largest ones.}, which we take to be located at $l=0$ and $l=L$, respectively, and apply the Cary-Shasharina theorem (\ref{CS}), we obtain
    \bn \frac{\partial}{\partial \alpha} \left[u(\psi,\alpha,L) - u(\psi,\alpha,0) \right] = 0. 
    \label{condition on u}
    \en
If $I(\psi)=0$, this condition not only implies that $ u(\psi,\alpha,L) - u(\psi,\alpha,0)$ is a constant independent of $\alpha$, but also that $u(\psi,\alpha,0) = 0$. This can be seen as follows. If we express $u$ in ordinary Boozer coordinates, $u = U(\psi,\theta,\varphi)$, where $\varphi = 2 \pi n/N_p$ at the maxima of the field strength, then $U(\psi,\theta,0)$ must vanish for {\em some} $\theta$, since the net toroidal current vanishes. Furthermore, the condition (\ref{condition on u}) and periodicity in $\varphi$ becomes
    $$ U(\psi,\theta + 2 \pi \iota / N_p,  2 \pi / N_p) = U(\psi,\theta + 2 \pi \iota / N_p,  0) = U(\psi,\theta,0),  $$
If the rotational transform $\iota$ is irrational, it follows that $U(\psi,\theta,0) = 0$ for all $\theta$. In other words, the parallel current vanishes at the boundary of each field period, $\varphi = 2 \pi n/N_p$, where the field strength reaches its maximum on each flux surface. The streamlines of $\bf J$ therefore cannot extend around the torus in the toroidal direction, but must `turn around' in each period of the stellarator \citep{Subbotin-2006,helander2014}. 

\begin{figure}
    \centering
    \includegraphics[width=0.5\linewidth]{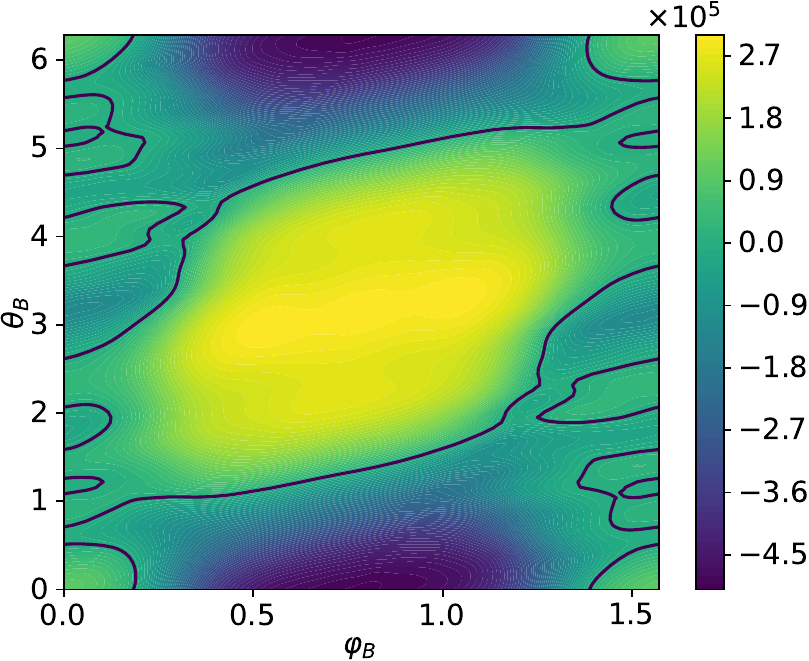}
    \caption{$J_\parallel$ ($\mathrm{A/m^2}$) on the $s=0.5$ flux surface of the "elongated QI" configuration from \citet{Goodman2024}, which has four field periods. Here $s$ is the normalized toroidal flux, with $s=0$ on axis and $s=1$ on the boundary, and $(\theta,\varphi)$ are the Boozer angles. The black curves show where $J_\parallel$ vanishes.}
    \label{fig:j_par}
\end{figure}

To give an example of what this looks like in a well-optimized QI field, we show a plot of $J_\parallel$ on a flux surface in the "elongated QI" configuration from \citet{Goodman2024} in Figure \ref{fig:j_par}. This configuration, which has four field periods, was optimized for good QI quality at finite beta, and we use it without modification. As can be seen, $J_\parallel$ reaches the its highest magnitude about half way through the period and its lowest magnitude close to $\varphi = 0$ and $\varphi = 2\pi/N_p = \pi/2$. Note that the black curves, which mark where $J_\| = 0$, stay close to these values of $\varphi$, except for those two that separate the positive and negative regions of maximum magnitude. In a perfect QI field, the black curves would exactly coincide with $\varphi = 0$ and $\varphi = 2\pi/N_p$, but imperfections in the quasi-isodynamicity, which are inevitably present in any configuration, lead to the observed deviations.

\section{Shafranov shift}

\subsection{Perturbed magnetic field}

In order to investigate the Shafranov shift, we consider the effect of the addition of a small amount of plasma pressure to an otherwise zero-beta MHD equilibrium. For simplicity, we consider this question within the framework of reduced magnetohydrodynamics \citep{Strauss-1997,Zocco-2021} and assume that the net toroidal current vanishes. In other words, the magnetic field is written as ${\bf B} = {\bf B}_0 + \delta {\bf B}$, where the unperturbed equilibrium corresponds to $\beta=0$ and can thus be written as
	\bn {\bf B}_0 = \nabla \psi_0 \times \nabla \theta + \nabla \varphi \times \nabla \chi_0 = G \nabla \varphi, 
	\label{B in magnetic coordinates}
	\en
with constant $G$. Here and in the following $(\psi,\theta,\varphi)$ denotes Boozer coordinates associated with the unperturbed field, and for simplicity we have taken the net toroidal current to vanish. 

The central assumption of RMHD is that there are two disparate scales on which the unperturbued and perturbed fields vary. The unperturbed magnetic field ${\bf B}_0$ is assumed to vary on a length scale $L_\|$ in all directions whereas the perturbation $\delta {\bf B}_0$ is taken to vary on the much shorter scale $L_\perp$ in the directions perpendicular to ${\bf B}_0$ whereas it only varies on the longer scale $L_\|$ along ${\bf B}_0$. In a stellarator, $L_\perp$ is thus identified with the minor radius $a$ and $L_\|$ with the major radius $R$.\footnote{In an arbitrarily shaped stellarator, $R$ can be defined as the length of the magnetic axis divided by $2 \pi$.} It follows that the gradients of the poloidal and toroidal angles scale as
    $$ |\nabla \theta| \sim \frac{1}{L_\perp}, \qquad |\nabla \varphi| \sim \frac{1}{L_\|}, $$
and the vectors $\nabla \theta$ and $\nabla \varphi$ are almost perpendicular since ${\bf B} = G \nabla \varphi$ and 
    $$ {\bf B} \cdot \nabla \theta = \iota {\bf B} \cdot \nabla \varphi \sim \frac{B}{L_\|}$$
imply $\nabla \theta \cdot \nabla \varphi \sim \epsilon |\nabla \theta ||\nabla \varphi|$ with $\epsilon = L_\perp / L_\| \ll 1$. The rotational transform $\iota = d\chi_0/d\psi_0$ is treated as a number of order unity, and the poloidal and toroidal fluxes are thus comparable, $\psi_0 \sim \chi_0 $, 
although the poloidal field is much smaller than the toroidal one. 

Turning to the MHD equations, we first note that the equilibrium condition 
	$$ \nabla \left( \mu_0 p + \frac{B^2}{2} \right) - {\bf B} \cdot \nabla {\bf B} = 0, $$
becomes, to leading order in $\epsilon$,
	$$ \nabla_\perp ({\bf B}_0 \cdot \delta {\bf B} + \mu_0 p) = 0 $$ 
and thus implies that $ \delta {\bf B} = \delta {\bf B}_\perp + {\bf b} \delta B_\|$, with ${\bf b} = {\bf B}_0 / B_0$ and 
	\bn \frac{\delta B_\|}{B_0} = - \frac{\mu_0 p}{B_0^2} = - \frac{\beta}{2}. 
    \label{delta Bpar}
    \en
Furthermore,  the relation $\nabla \cdot \delta {\bf B}_\perp = - \nabla \cdot ( {\bf b} \delta B_\|) \sim \epsilon \delta B_\perp / L_\perp$ implies that we can write
	\bn \delta {\bf B}_\perp = \nabla \varphi \times \nabla \delta \lambda + O(\epsilon^2 B_0) 
    \label{delta B-perp1}
    \en
for some function $\delta \lambda(\psi,\theta,\varphi) \ll \chi_0$. (All perturbations are marked with a $\delta$.) This stream function of the perpendicular field can be related to the poloidal and toroidal flux perturbations by noting that $ \delta {\bf B}$ can also be written as
    \bn \delta {\bf B} = \nabla \delta \psi \times \nabla \theta + \nabla \varphi \times \nabla \delta \chi  
    + \nabla \psi \times \nabla \delta k  
    \label{delta B}
    \en
for some functions $\delta \psi$, $\delta \chi$ and $\delta k$.\footnote{Note that $(\psi,\theta,\varphi)$ denote Boozer coordinates of the unperturbed equilibrium, which may differ slightly from those of the perturbed one.} Hence it follows that, to leading order,
    \bn \delta {\bf B}_\perp \times \nabla \varphi =  |\nabla \varphi|^2 \nabla \delta \chi 
    - (\nabla \theta \cdot \nabla \varphi) \nabla \delta \psi = |\nabla \varphi|^2 \left( \nabla \delta \chi - \iota \nabla \delta \psi \right), 
    \label{delta B-perp2}
    \en
where we have used the fact that, for any perturbation $\delta f$, the derivative along ${\bf B}_0$ is relatively small, 
    $$ {\bf B}_0 \cdot \nabla \delta f \ll B_0 |\nabla \delta f |. $$
According to (\ref{delta Bpar}), the toroidal-flux perturbation is of order $\delta \psi = O(\beta \psi)$, but the perturbation in poloidal flux is much larger, $\delta \chi = O(\delta \psi / \epsilon)$, since the poloidal and toroial field perturbations are comparable, $\delta B_\perp \sim \delta B_\|$. It follows that the last term in (\ref{delta B-perp2}) is relatively small, and from (\ref{delta B-perp1}) and (\ref{delta B-perp2}) we thus conclude that $\delta \lambda$ is approximately equal to the poloidal-flux perturbation, $\delta \lambda \simeq \delta \chi$.

In conclusion, then, the addition of a small amount of pressure, $\beta \ll \epsilon$, to a vacuum field (\ref{B in magnetic coordinates}) causes the poloidal flux to change from $\chi_0$ to $\chi = \chi_0 + \delta \chi$, where $\delta \chi/\chi_0 = O(\beta/\epsilon)$. This perturbation leads to a slight movement of the plasma column, which will be quantified in the next subsection by making use of the fact that the parallel current is related to $\delta \chi$ by
	\bn \mu_0 J_\|
	 = {\bf b} \cdot ( \nabla \times \delta {\bf B}_\perp )
 	\simeq \frac{B}{G} \nabla^2 \delta \chi, 
 	\label{Ampere}
 	\en
to leading order in $\epsilon$. 

\subsection{Average Shafranov shift}

The movement and the deformation of the flux surfaces constitute the Shafranov shift, which depends both on the pressure-induced currents inside the plasma and on the boundary condition at the plasma edge. The Shafranov shift is therefore different in fixed-boundary and free-boundary equilibria. In the former, the fixed boundary remains a flux surface as pressure is added whereas in free-boundary equilibria the plasma boundary will in general be deformed. We focus on the former case, where $\chi=0$ on the fixed boundary. 

\begin{figure}
    \centering
    \includegraphics[width=0.5\linewidth]{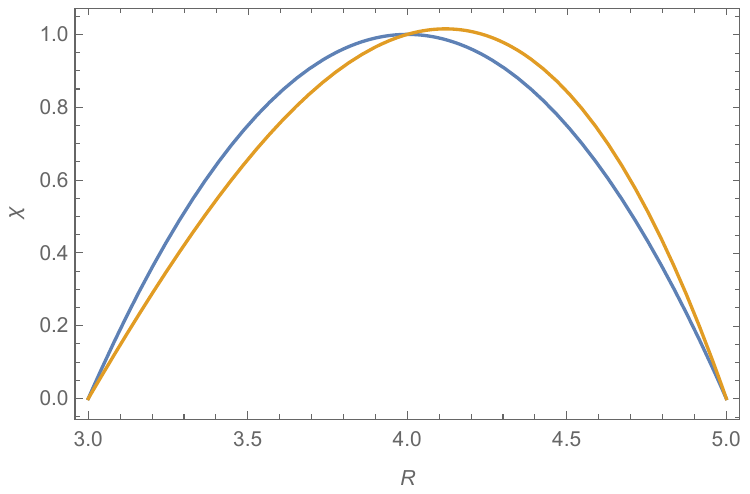}
    \caption{Schematic figure illustrating how the poloidal flux $\chi$ may vary along a long a ray (here in terms of the major radius $R$) in a fixed-boundary equilibrium without (blue) and with (orange) plasma pressure. The poloidal flux vanishes at the boundary and peaks on the magnetic axis. With increasing pressure, the magnetic axis moves outward, and the spatial distribution of $\chi$ changes accordingly. }
    \label{fig:flux}
\end{figure}

In order to quantify the average Shafranov shift, we introduce the functional
	$$ S[\chi] = \frac{1}{X} \int_P \chi F dV , $$
	$$ X = \int_P \chi dV, $$
where $F({\bf r})$ is a suitably chosen function of the coordinates $\bf r$ and the integral is taken over the plasma volume $P$. $S[\chi]$ can be regarded as a $\chi$-weighted average of the function $F({\bf r})$ over the domain $P$. The Shafranov shift is the amount by which $S[\chi]$ changes due to the redistribution of poloidal flux when pressure is added, see Figure \ref{fig:flux}. To first order in $\delta \chi/\chi_0 \ll 1$, we have
	$$ \delta S = S[\chi_0 + \delta \chi] - S[\chi_0] = \frac{1}{X} \left( \int_P \delta \chi F dV 
	- S[\chi_0] \int_P \delta \chi dV \right), $$
where the second term in the brackets vanishes if either $F({\bf r})$ is chosen so that $S[\chi_0] = 0$ or the volume integral of $\delta \chi$ vanishes.\footnote{In the numerical examples below, we choose $F$ in such a way that $S[\chi_0] = 0$. If, for some reason, such a choice is undesirable, the volume integral of $\delta \chi$ can instead be made to vanish by adjusting the overall field strength in the finite-$\beta$ equilibrium appropriately.} In the following, we shall assume this to be the case. 

Different choices of $F$ correspond to different manifestations of the Shafranov shift. For instance, choosing $F = \delta({\bf r} - {\bf r}_0)$ is synonymous with declaring an interest in the value of the poloidal flux in the point ${\bf r}_0$. If this point lies on the magnetic axis of the vacuum field, then $\delta S$ measures how far the magnetic axis moves away from that point, due to the addition of pressure, in terms of poloidal flux. (If $F$ is chosen in this way, then $S[\chi_0] \ne 0$ and we must instead require that the volume integral of $\delta \chi$ vanishes.) Another choice of interest is 
	$$ F({\bf r}) = x({\bf r}), $$
where $x$ is a coordinate in the plane perpendicular to ${\bf B}_0$ that increases in a direction that is deemed interesting. $S[\chi]$ then measures a $\chi$-weighted average of $x$ over the domain, and $\delta S$ indicates how far the poloidal flux has moved, on average, in the $x$-direction. Whatever the exact definition of $F$, we shall normalise it in such a way that a Shafranov shift of order unity, $\delta S = O(1)$, corresponds to a displacement of the plasma column comparable to the minor radius $a$.

In order to derive a useful expression for $\delta S$, we introduce a function $H({\bf r})$ such that
	\bn \nabla^2 H = F 
    \label{H}
    \en
everywhere in the plasma volume $P$ and $H=0$ on the boundary. These conditions uniquely determine $H$, which encapsulates the necessary information about the shape of the plasma cross section and is discussed in Appendix A. Our measure of the Shafranov shift can now be written as
	\bn X \delta S = \int_P \delta \chi \nabla^2 H dV 
	= \int_P H \nabla^2 \delta \chi \; dV
	= \mu_0 G \int_P \frac{H J_\|}{B} \; dV, 
	\label{final SS}
	\en
where we have used (\ref{Ampere}) and the fact $\delta \chi$ vanishes on the boundary. Note that, in reduced MHD, the Shafranov shift is exclusively caused by the parallel (Pfirsch-Schlüter) current to leading order in $\epsilon$. It is assumed that the net toroidal current vanishes, on each flux surface, both before and after the addition of pressure. The case of a `flux-conserving stellarator' \citep{Freidberg} is thus not considered. 

A general, yet explicit, expression for the Shafranov shift in a stellarator without net toroidal current can be obtained by substituting (\ref{JPS-exact}) in (\ref{final SS}),
    \bn \delta S = \frac{\mu_0 G^2}{X} \sum_{m,n\ne 0}  \int_P  \frac{dp}{d\psi} \frac{h_{mn} H}{\iota - n/m} 
    e^{i(m \theta - n \varphi)} \; dV, 
    \label{General SS}
    \en
where all necessary information about the magnetic-field geometry is encapsulated in the quantities $X$, $h_{mn}$ and $H$. 

An equally general upper bound on $\delta S$ follows from (\ref{final SS}) and the Cauchy-Schwarz inequality,
	\bn \delta S \le \frac{\mu_0 G}{X} \left[ \int_P H^2 dV
	\int_P \left( \frac{J_\|}{B} \right)^2 dV \right]^{1/2}, 
	\label{upper bound}
	\en
where $J_\|$ is related to the pressure by (\ref{magnetic DE}). It is clear from (\ref{final SS}) and (\ref{upper bound}) that the Shafranov shift is particularly small in a magnetic field optimised for small Pfirsch-Schl{\"u}ter current \citep{Wobig}. 

Before closing this section, we note that there is no way of avoiding the Shafranov shift altogether by optimising the magnetic-field geometry. The average Shafranov shift (\ref{final SS}) does not vanish for all allowable functions $H$ unless the Pfirsch-Schl{\"u}ter current $J_\|$ vanishes everywhere. This happens only if the field lines are geodesics, i.e. if the magnetic field is isodynamic, which is impossible in practice \citep{helander2014}. In the context of axisymmetric fields, \cite{Bishop} considered the related question (without the reduced-MHD approximation) by asking whether the Grad-Shafranov equation has solutions with identically shaped flux surfaces but different pressure and current profiles. 

\section{Scalings}

From the results above, it is possible to deduce certain general properties of the Shafranov shift in various types of magnetic fields. This is the subject of the present section. The exact choice of the function $F$ will be left unspecified except that it is taken to be dimensionless and of order unity, which implies $H \sim a^2$ according to the definition (\ref{H}). 

\subsection{General geometry}

In Boozer coordinates, (\ref{magnetic DE}) becomes
    \bn \frac{\partial u}{\partial \varphi} + \iota \frac{\partial u}{\partial \theta} = - G  \frac{\partial B^{-2}}{\partial \theta} = f(\theta,\varphi), 
    \label{u-eq}
    \en
where again $u = J_\| / (Bp'(\psi))$, and since $\partial B/\partial \theta = O(\epsilon)$ in the RMHD ordering, it follows that $ u \sim G \epsilon / B^2$.
This general scaling shows how $u$ depends on $\epsilon$ when $\iota$ is order unity, but it is modified if $\iota \gg 1$, in which case an additional factor $\iota$ appears in the denominator,
        \bn u \sim \frac{G \epsilon}{\iota B^2}. 
        \label{u-scaling}
        \en
This is perhaps obvious already from (\ref{u-eq}) but can also be verified by inspecting the solution
    \bn u(\theta,\varphi) = \int_{\varphi_0}^\varphi f\left[ \theta + \iota ( \varphi' - \varphi), \varphi' \right] d\varphi', 
    \label{u-integral}
    \en
where $\varphi_0(\theta,\varphi)$ is chosen so that $u[\theta + \iota ( \varphi_0 - \varphi),\varphi_0] = 0$. In the limit $\iota \gg 1$, the integrand $g(\varphi') = f\left[ \theta + \iota ( \varphi' - \varphi), \varphi' \right]$
becomes almost periodic with a period of $2 \pi / \iota$ since
    $$ g(\varphi' + 2\pi / \iota)  -  g(\varphi') = O\left( \iota^{-1} \right). $$
More importantly, its integral over one period is of order
    $$ \int_{\varphi_0 + 2 \pi n/\iota}^{\varphi_0 + 2 \pi (n+1)/\iota} g(\varphi') d\varphi' = O\left(\iota^{-2}\right), $$
since, by the definition of $f(\theta,\varphi)$
    $$ \int_0^{2\pi} f(\theta, \varphi) d\theta = 0. $$
Hence we conclude that $u \sim f/\iota$, and (\ref{u-scaling}) follows if $\iota$ is either comparable to, or larger than, unity. If we further estimate, within factors of order unity, $G \sim BR$, $\chi \sim \iota \psi \sim \iota B a^2$, we arrive at a universal scaling of the average Shafranov shift (\ref{final SS}) 
    \bn \delta S \sim \frac{\beta}{\epsilon \iota^2}. 
    \label{general scaling}
    \en
The equilibrium pressure limit is thus expected to scale as (\ref{classical}), indicating that Spitzer's result is of general validity to the present level of accuracy. As long as the magnetic field follows the RMHD ordering, this is how the equilibrium beta limit scales with $\epsilon$ and (more approximately) with $\iota$. 

Slightly more precise results can be derived for quasisymmetric and quasi-isodynamic stellarators, to which we now turn our attention. 

\subsection{Quasisymmetric stellarators}

In a quasisymmetric stellarator without net toroidal current, $I=0$, Eqs.~(\ref{QS JPS}) and (\ref{final SS}) give the following explicit formula for $\delta S$,
		\bn \delta S_{\rm QS}  = \frac{\mu_0 G^2}{X} \int_P \frac{p'(\psi)}{N-\iota} 
		 \lang H \left( \frac{1}{B^2} - \frac{1}{\lang B^2 \rang} \right) \rang dV. 
		\label{QS SS}
		\en
According to the ordering underlying reduced MHD, the magnetic field strength $B$ only varies on the long length scale $L_\|$, which is taken to be comparable with the circumference of the torus. In a quasi-axisymmetric (QA) or quasi-helically (QH) symmetric stellarator, $B$ is constant along the magnetic axis and can therefore only vary by a relative amount of order $\epsilon = L_\perp/L_\| \ll 1$ in the entire plasma volume. Again using $G \sim BR$ and $\chi \sim \iota B a^2$ gives
    $$ \frac{J_{\|}^{\rm PS}}{B} = \frac{G}{N-\iota} \frac{dp}{d\psi} \left( \frac{1}{B^2} - \frac{1}{\lang B^2 \rang} \right)
    \sim \frac{RB}{N - \iota} \frac{p}{a^2B} \frac{\epsilon}{B^2} = \frac{p}{(N-\iota) aB^2}, $$
within factors of order unity. It follows that the Shafranov shift (\ref{QS SS}) in a quasisymmetric stellarator is of order
    \bn \delta S_{\rm QS} \sim \frac{\beta}{|N-\iota|\iota \epsilon}, 
    \label{SQS}
    \en
where we have again crudely estimated the poloidal flux by $\chi \sim \iota \psi$. The equilibrium pressure limit is thus expected to scale as
    \bn \beta_{\rm QS} \sim \left| \frac{N}{\iota} - 1 \right| \epsilon \iota^2. 
    \label{QS beta limit}
    \en
These scalings agree with with the classical-stellarator result (\ref{classical}) but contain an additional dependence on $N / \iota$.

\subsection{Quasi-isodynamic stellarators}

In a QI stellarator, the Shafranov shift must of course follow the approximate general scaling (\ref{general scaling}), but it is also possible possible derive a quantitatively precise result. As we have already seen, the Pfirsch-Schl{\"u}ter current in a QI stellarator without net toroidal current satisfies (\ref{eq for u}), but unlike the quasisymmetric case, there is no closed form for the solution of this equation. 
It is however possible to use (\ref{upper bound}) to derive an upper bound on the Shafranov shift. We start by noting that  
	$$ \lang u^2 \rang = \frac{1}{V'(\psi)} \oint d\alpha \int_0^Y u^2 dy, $$
where $dy = dl/B$ and the integral
	$$ Y(\psi) = \int_0^L \frac{dl}{B}, $$
is performed along $\bf B$ over one period of the device, $\varphi \in [0,2\pi /N_p]$. By using the Poincaré inequality, 
    $$ \int_0^Y u^2 dy \le \left( \frac{Y}{\pi} \right)^2 \int_0^Y \left( \frac{\partial u}{\partial y} 
    \right)^2 dy, $$
for functions such that $u(0) = u(Y) = 0$, we conclude from (\ref{eq for u}) that
    $$ \lang u^2 \rang \le \frac{4Y^2}{\pi^2 V'} 
    \oint d\alpha \int_0^L \left( \frac{\partial \ln B}{\partial \alpha} \right)^2 
    \frac{dl}{B}. $$
Since $V'(\psi) = 2 \pi Y(\psi)$, this result can be written as
    $$ \lang \left( \frac{J_\|}{B} \right)^2 \rang 
    \le \frac{1}{\pi^4} 
    \lang \left( p' V' \frac{\partial \ln B}{\partial \alpha} \right)^2 \rang,
    $$
which can be substituted into (\ref{upper bound}) to give an upper bound on the Shafranov shift in QI stellarators without net toroidal current,
	\bn \delta S_{\rm QI} \le \frac{\mu_0 G}{\pi^2 X} \left[ \int_P H^2 dV
	\int_P 
    \left(p' V' \frac{\partial \ln B}{\partial \alpha} \right)^2 dV \right]^{1/2}.
	\label{QI upper bound}
	\en
Ignoring factors of order unity in the estimates $ V'(\psi) \sim {2 \pi R}/({N_pB})$,  $p'(\psi) \sim p/(a^2 B)$ and
    \bn \frac{\partial \ln B}{\partial \alpha} \sim \epsilon, 
    \label{dB/da}
    \en
 gives the scaling
    \bn \delta S_{\rm QI} \simlt \frac{\beta}{N_p\iota \epsilon}, 
    \label{SQI}
    \en
thus suggesting a beta limit 
    $$ \beta_{\rm QI} \sim \frac{N_p}{\iota} \epsilon \iota^2. $$
Apart from the dependence on $N_p/\iota$, this scaling agrees with that for classical stellarators (\ref{classical}) and quasisymmetric ones (\ref{SQS}). It should be noted that, in many cases, (\ref{dB/da}) is likely to be an overestimate of the poloidal magnetic-field-strength variation since the field strength may vary by less than $O(\epsilon)$ in the poloidal direction of a QI stellarator. This variation vanishes identically on the contour $B = B_{\rm max}(\psi)$ and can be made quite small elsewhere by targetted optimisation of the magnetic-field geometry, which then results in a correspondingly smaller Shafranov shift.

\section{Numerical results}

In order to explore the Shafranov shift numerically, a series of magnetic equilibria based on the following configurations were computed: 
\begin{enumerate}
\item the Wendelstein 7-X high-mirror configuration (KJM252);
\item one of the Stable Quasi-Isodynamic Designs (SQuIDs) of \cite{Goodman2024};
\item a `low-$\iota$ SQuID', which was optimised in the same way as (ii) but for a lower average $\iota$;
\item a `precise QA configuration' of \cite{LandremanPaul};
\item a `precise QH configuration' of \cite{LandremanPaul}.
\end{enumerate}

\begin{figure}
    \centering
    \includegraphics[height=3.6cm]{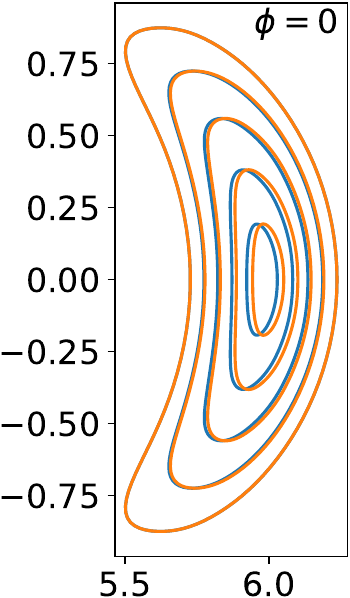}\hfill\includegraphics[height=4.15cm]{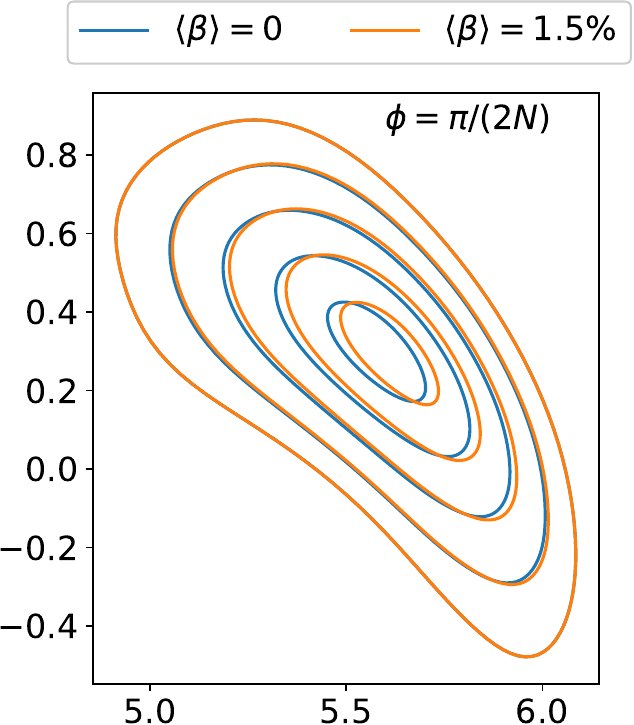}\hspace{1cm}\includegraphics[height=3.6cm]{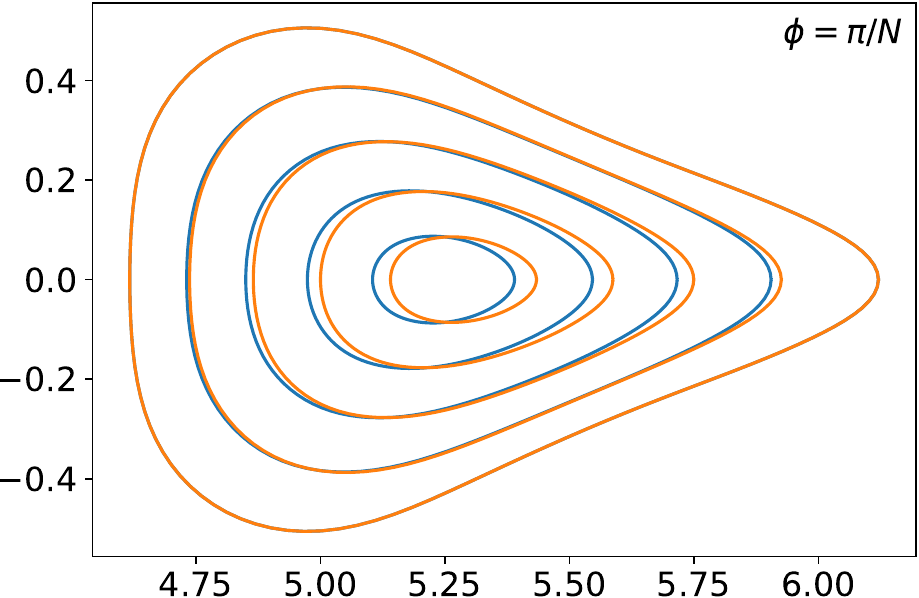}
    \caption{Flux surface plots in Wendelstein 7-X (no changes to boundary coefficients) at three different cross sections, shown for zero and finite beta.}
    \label{fig:w7x_cross}
\end{figure}

\begin{figure}
    \centering
    \includegraphics[height=3.6cm]{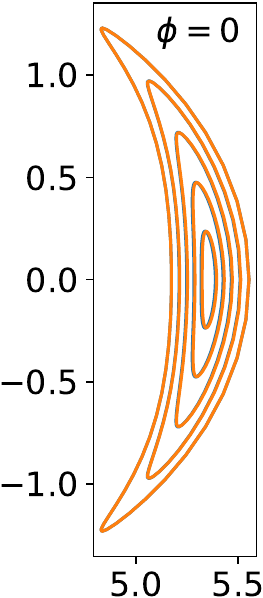}\hfill\includegraphics[height=4.15cm]{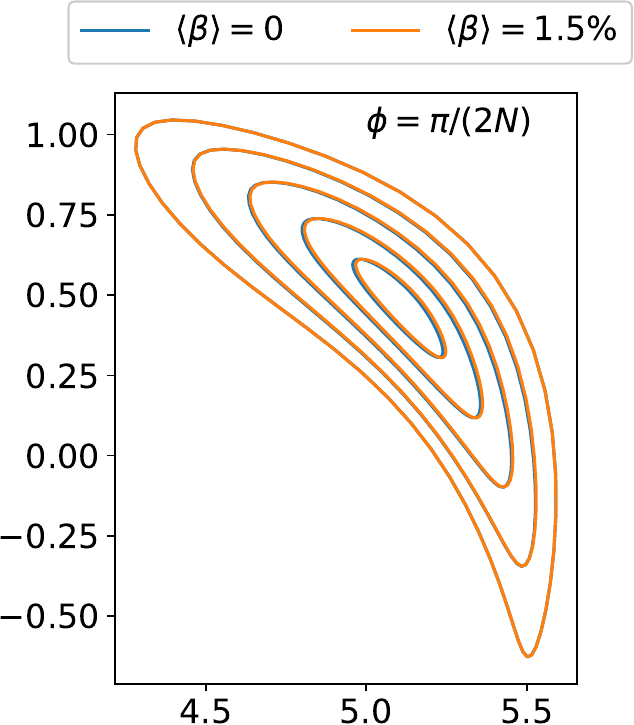}\hspace{1cm}\includegraphics[height=3.6cm]{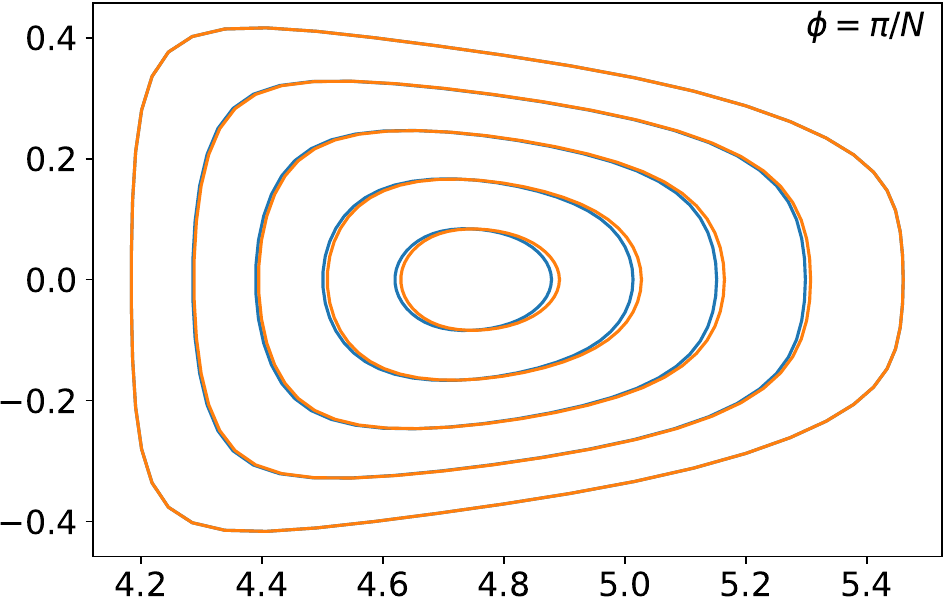}
    \caption{Flux surface plots in the SQuID (no changes to boundary coefficients) at three different cross sections, shown for zero and finite beta.}
    \label{fig:squid_cross}
\end{figure}

\begin{figure}
    \centering
    \includegraphics[height=3.6cm]{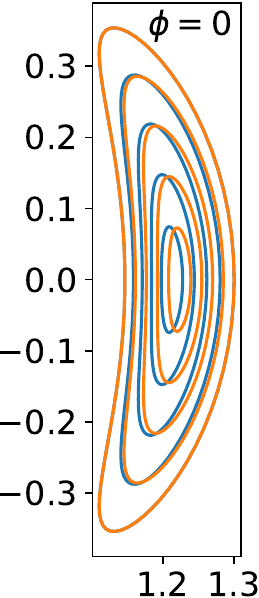}\hfill\includegraphics[height=4.15cm]{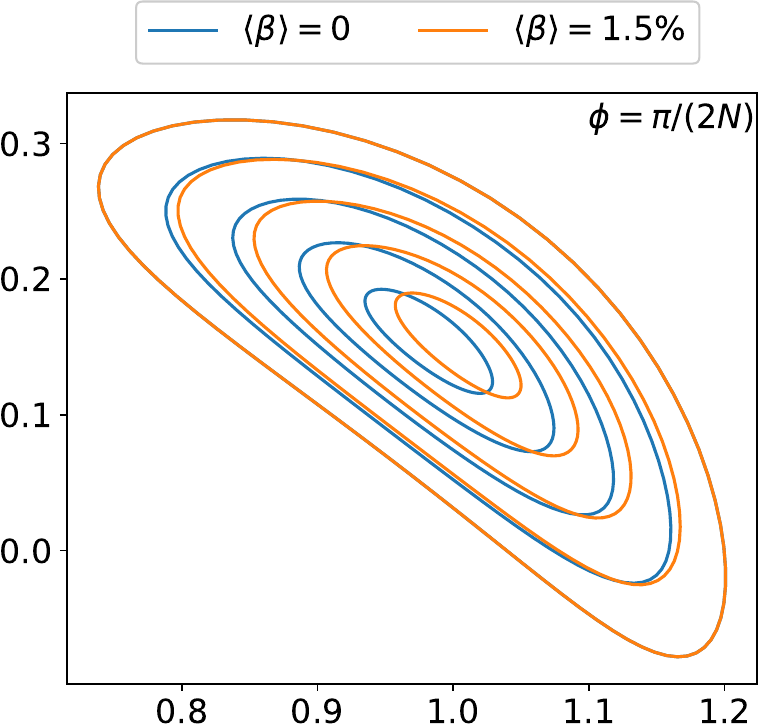}\hspace{0.25cm}\includegraphics[height=3.6cm]{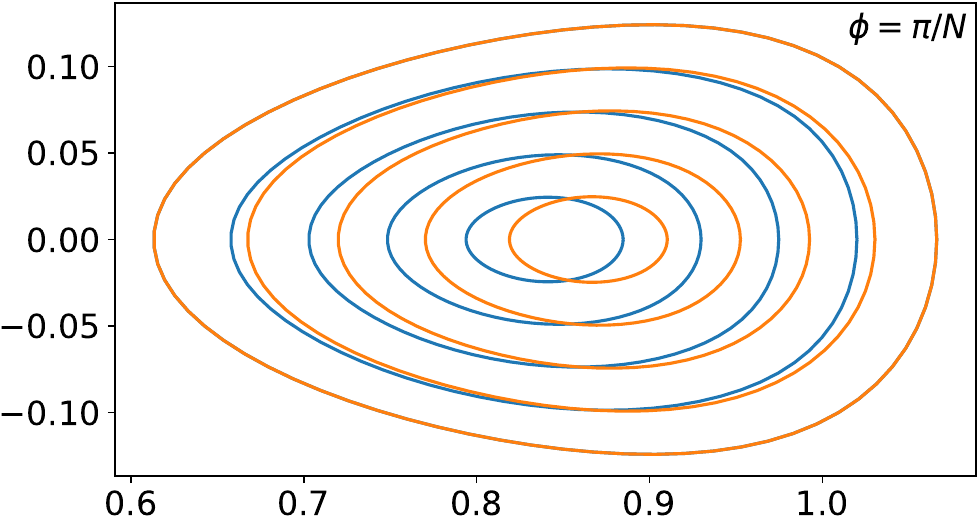}
    \caption{Flux surface plots in the Precise QA (no changes to boundary coefficients) at three different cross sections, shown for zero and finite beta.}
    \label{fig:pqa_cross}
\end{figure}

\begin{figure}
    \centering
    \includegraphics[height=3.6cm]{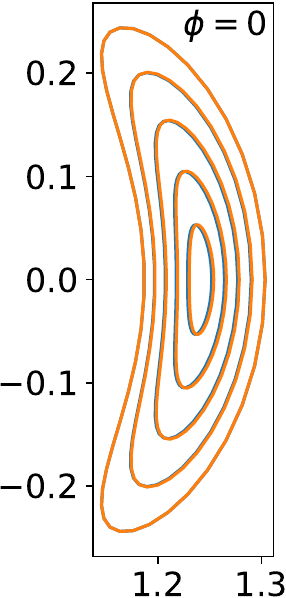}\hspace{0.25cm}\includegraphics[height=4.15cm]{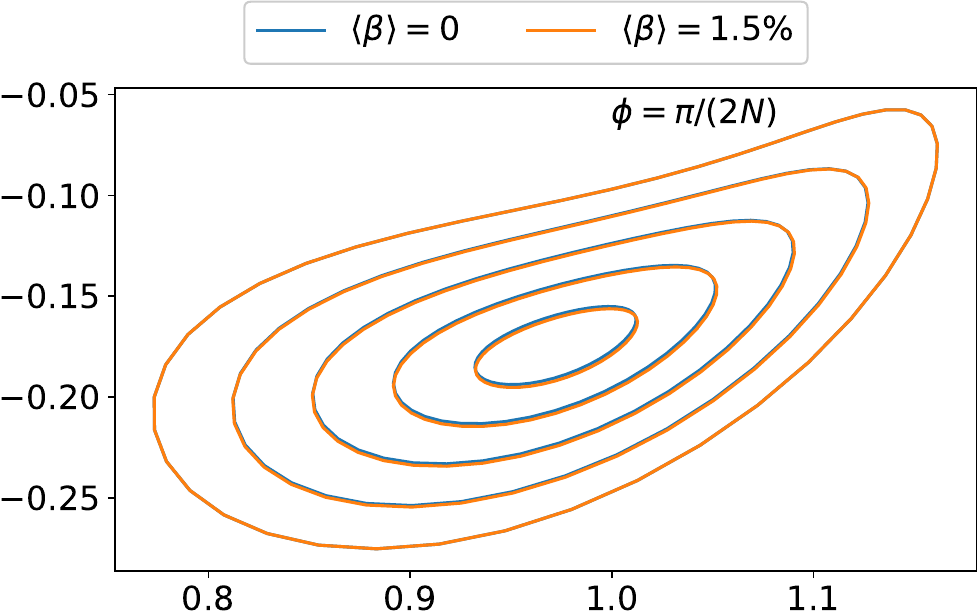}\\
    \vspace{0.1cm}
    \includegraphics[height=3.6cm]{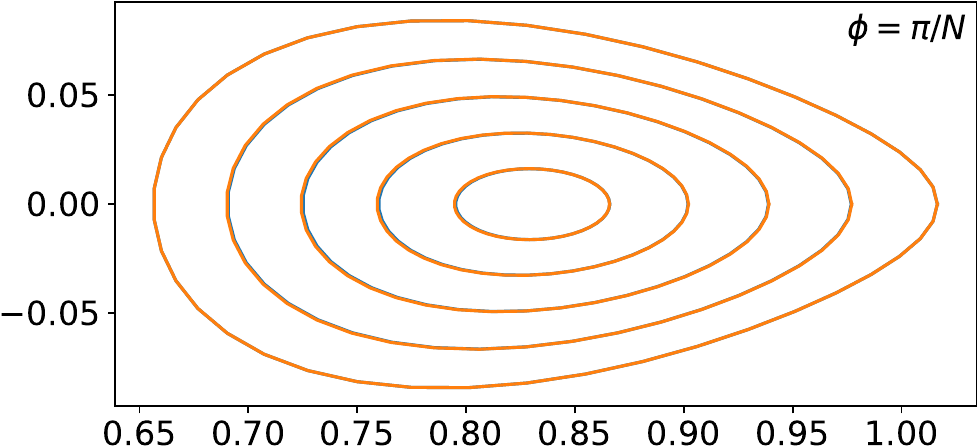}
    \caption{Flux surface plots in the Precise QH (no changes to boundary coefficients) at three different cross sections, shown for zero and finite beta.}
    \label{fig:pqh_cross}
\end{figure}

First of all, fixed-boundary equilibria with volume-averaged beta values of $\langle \beta \rangle = 0$ and $\langle \beta \rangle = 0.015$ were computed using the VMEC code \citep{VMEC}. In the poloidal cross sections shown in Figures \ref{fig:w7x_cross} -  \ref{fig:pqh_cross}, the outward shift of the flux surfaces can clearly be seen. The shift is smallest in the QH and SQuID configurations, and largest in the QA one. Again, we stress that the net toroidal current is neglected although the bootstrap current may, depending on plasma parameters, be important in QA and QH configuratioons.

Next, these equilibria were modified in a manner that will now be described. In stellarator design, the shape of the plasma boundary is usually prescribed in terms of Fourier series in cylindrical coordinates $(R,\phi,z)$,
    $$ R(\theta,\varphi) = \sum_{m,n} R_{mn} \cos\left(m \theta - nN_p \phi \right), $$
    $$ z(\theta,\varphi) = \sum_{m,n} z_{mn} \sin\left (m \theta - n N_p \phi \right), $$
where, for simplicity, we have only included stellarator-symmetric terms. This representation is not unique and can be improved \citep{Henneberg_Helander_Drevlak_2021} but provides a simple way to represent the boundary shape, which can be varied by adjusting the coefficients $R_{mn}$ and $z_{mn}$.

\begin{figure}
    \centering
    \includegraphics[width=0.48\linewidth]{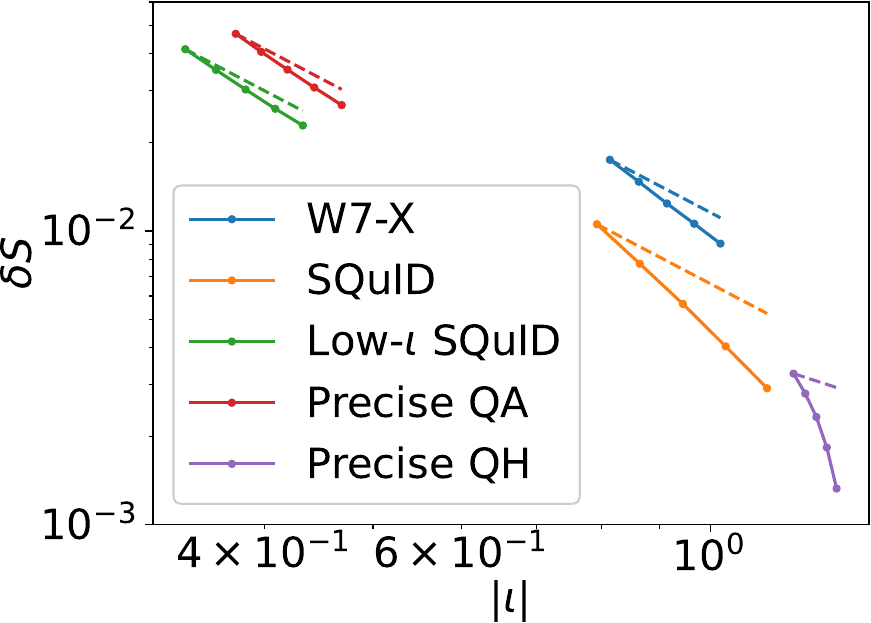}
    \includegraphics[width=0.48\linewidth]{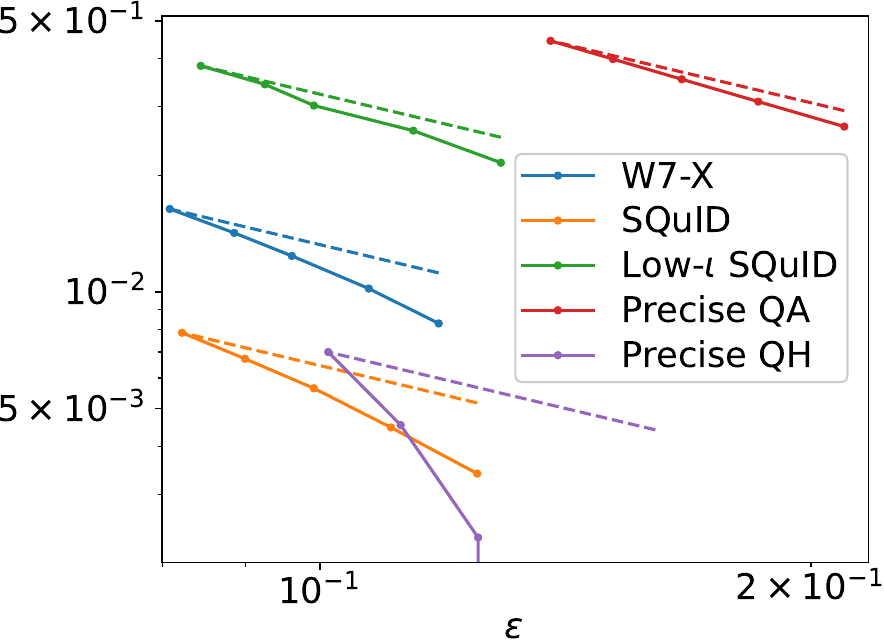}
    \caption{The values of $\delta S$ for all five devices as a function of $1/\iota$ with $\epsilon$ fixed (left), and as a function of $1/\epsilon$ with $\iota$ fixed (right). Trend lines showing the expected $1/\iota^2$ and $1/\epsilon$ scalings, respectively, are dashed.}
    \label{fig:compare_all}
\end{figure}

In order to investigate the dependence on $\iota$, the rotational transform was varied while the aspect ratio was kept fixed. In the configurations (i) - (iii) listed above, this was accomplished by multiplying all boundary coefficients $z_{mn}$ that have $n=\pm 1$ by a constant factor $k$, which changes the vertical elongation of the poloidal cross section of the flux surfaces. Close to the magnetic axis, the cross section is elliptical and rotates as one goes toroidally around the torus, thus producing rotational transform \citep{Mercier_1964}. The elongation of the rotating ellipse in Mercier's formula for the rotational transform either increases for $k>1$ or decreases for $k<1$, and in this way $\iota$ is varied on the magnetic axis. The coefficient $R_{00}$ is then adjusted in such a way that the aspect ratio remains the same as when $k=1$. In the QH configuration listed under (iv) above, most of the rotational transform comes from torsion of the magnetic axis\footnote{Indeed, the on-axis effective rotational transform, $\iota_N = \iota - N$, is 2.755 in the precise QH configuration. Using pyQSC \citep{pyQSC}, we determined the contribution from axis torsion to be 2.042 and from ellipticity to be 0.713.}, which was therefore varied instead of the ellipticity, by multiplying the coefficients $z_{0n}$ by a factor $k$.\footnote{It is important to note that the geometry of the magnetic field varies in more than one way by these operations. As a result, the Shafranov shift will change not merely because of the variation in $\iota$ but also due to other changes, which in our formalism are encapsulated in the function $H$. As a consequence of such effects, if $\iota$ is varied in QH configuration by the same method as in the other ones, $\delta S$ actually varies in the opposite way to that shown in the left panel of Figure \ref{fig:compare_all}.}  The resulting $\delta S$ in a series of magnetic fields corresponding to the values $k = \{0.8, 0.9, 1, 1.1, 1.2 \}$ are displayed in Figure \ref{fig:compare_all}. In the definition of $\delta S$, the choices
    $$F = \frac{R - R_0}{a}, $$
    $$ R_0 = \int\chi_0 RdV \bigg\slash \int\chi_0 dV, $$
were made, where $R$ is the cylindrical radial coordinate and $a$ the minor radius as given by VMEC. This choice ensures $S[\chi_0] = 0$, as required. 

The Shafranov shift was also computed in a second set of modified equilibria, shown in the right panel of Figure \ref{fig:compare_all}, where the rotational transform on axis was held fixed while the aspect ratio was varied. This time, the coefficient $R_{00}$ was varied to change the aspect ratio while $k$ as adjusted to preserve the on-axis rotational transform $\iota$. The values $R_{00} = \{4.55, 5.0, 5.55, 6.0, 6.55\} \mathrm{m}$ were chosen for Wendelstein 7-X, $R_{00} = \{4.0, 4.5, 5.0, 5.5, 6.0\}\mathrm{m}$ for the SQuID, $R_{00} = \{13.82, 15.55, 17.27, 19.00, 20.73\}\mathrm{m}$ for the low-$\iota$ SQuID, and $R_{00} = \{0.8,0.9,1.0,1.1,1.2\} \mathrm{m}$ for precise QA and QH configurations, where $R_{00} = 5.55\mathrm{m}$, $R_{00} = 5.0\mathrm{m}$, $R_{00} = 17.27\mathrm{m}$ and $R_{00} = 1.0\mathrm{m}$ are the values in the original cases.

Modifying optimised equilibria in this simple way generally violates omnigenity: there is no guarantee that the modified fields remain quasi-symmetric or quasi-isodynamic, but they should remain approximately so if $k$ is close to unity. This question is discussed in greater detail in Appendix B. An alternative procedure would be re-optimise the equilibria for each value of $k\ne 1$, but such an exercise goes beyond the scope of this paper. 

The results in Figure \ref{fig:compare_all} all show that the Shafranov shift is a decreasing function of $\epsilon$ and $\iota$. The dependence on $\iota$ is stronger than that on $\epsilon$, as expected from the general scaling (\ref{general scaling}), which however underpredicts the dependence on both parameters. The lack of precise agreement can probably be attributed to the fact that, when the MHD equilibria were modified so as to vary $\epsilon$ and $\iota$, many other aspects of the geometry also changed. For instance, when the elongation is varied, the function $H$ changes, which is not accounted for in (\ref{general scaling}). This shortcoming of the scaling law is inevitable, since the parameters $\epsilon$ and $\iota$ only capture a small fraction of all the geometric freedom inherent in stellarator design. This freedom can be used to minimise the Shafranov shift, as is apparent from the fact that the QI and QH examples shown have much smaller shifts than the other configurations, as expected from the theory. 

Our QH configurations show particularly poor agreement with the scaling (\ref{general scaling}), which has to do with the fact that the factor $1/B^2 - 1 / \lang B^2 \rang$ appearing in (\ref{QS SS}) varies much more in the scans than predicted by the simple estimate $\epsilon$ used in the scaling. In addition, there is a general complication having to do with the fact that the Shafranov shift displaces the magnetic axis in different directions in different toroidal locations of QH stellarators \citep{Rodriguez_2023}. The quantity $\delta S$ plotted in Figure \ref{fig:compare_all} is a weighted average, over the entire plasma volume, of the displacement in the direction of the major radius. In some parts of the plasma, the displacement is negative (i.e. in the direction $-\nabla R$), and the fraction of such parts varies with $\epsilon $ and $\iota$. This circumstance, which is not reflected in the scaling, actually causes $\delta S$ to become negative if $\epsilon$ is increased beyond the displayed points. 

\section{Conclusions}

Most previous treatments of the Shafranov shift have focused on special cases, usually by calculating the shift in specific geometries, either analytically or numerically. By contrast, we have introduced a measure of the average Shafranov shift, which can be calculated in {\em any} stellarator with large aspect ratio. The result is given by (\ref{General SS}) and a generally valid upper bound by (\ref{upper bound}). These results enable a general discussion of the Shafranov shift in entire classes of stellarators rather than individual cases. 

The classical scaling (\ref{general scaling}) is found to hold generally for any stellarator with large aspect ratio and rotational transform, $\epsilon \ll 1 \ll \iota$. However, $\iota$ is rarely large in practice, and the Shafranov shift depends on additional aspects of the magnetic geometry. For instance, it depends on $N$ in quasisymmetric stellarators and $N_p$ in quasi-isodynamic ones, which can be utilised to make the Shafranov shift small. In QI devices, the Shafranov shift can also be minimised by reducing the poloidal variation of the magnetic field strength. As a result, recently optimised QI designs have a Shafranov that is only about half as large as that in W7-X, and it is yet smaller in QH configurations. 

For simplicity, we have limited the discussion to stellarators with small toroidal current. However, the bootstrap current could play an important role in quasisymmetric stellarators, where it can significantly modify the rotational transform, which, in turn, affects the Shafranov shift.

\section*{Declaration of interests}

The authors report no conflict of interest.

\appendix
\section{The function $H$}

The function $H$ defined in (\ref{H}) summarises the information about flux-surface shaping necessary for calculating the average Shafranov shift. Thanks to the orderings of reduced MHD, only gradients perpendicular to the magnetic field need to be included in the Laplace operator, and (\ref{H}) is thus a Poisson equation in two dimensions that determines $H$ separately in each plasma cross section defined by a cut perpendicular to the magnetic axis. If $(x,y)$ denote Cartesian coordinates perpendicular to the magnetic axis in such a cross section, (\ref{H}) is thus simply
    $$ \frac{\partial^2 H}{\partial x^2} + \frac{\partial^2 H}{\partial y^2} = F(x,y). $$
For instance, in the simple case that $F = x$ and the cross section is a circle of radius $a$, the solution becomes 
$$ H = \frac{x}{8} (x^2 + y^2 -a^2). $$

For a generally shaped cross section, $H$ can be calculated by writing $H=H_0 + H_1$, where $H_0$ is any solution to the equation 
    $$ \nabla^2 H_0 = F, $$
not necessarily satisfying the boundary condition $H_0=0$, then 
    $$ \nabla^2 H_1 = 0 $$
and $H_1=-H_0$ on the boundary. In a circular domain, the solution to this equation is given by Poisson's integral formula, and the solution in a differently shaped domain can be constructed by conformal mapping. 

\section{Non-omnigenity of modified equilibria}

As remarked in the main text, if the original magnetic equilibria are well optimised in the sense that they are exactly omnigenous for $k=1$, they will in general not be omnigenous if $k\ne 1$. However if $k-1 \ll 1$, the degree of non-omnigenity should be quadratic at the leading order of the Taylor expansion in this quantity. By constrast, the change in Shafranov shift can be either positive or negative, and therefore its Taylor series has a linear term in $k-1$. It is thus to be expected that $\delta S$ varies more rapidly with $k$ than the degree of omnigenity for small values of $k-1$.

\begin{figure}
    \centering
    \includegraphics[width=0.45\linewidth]{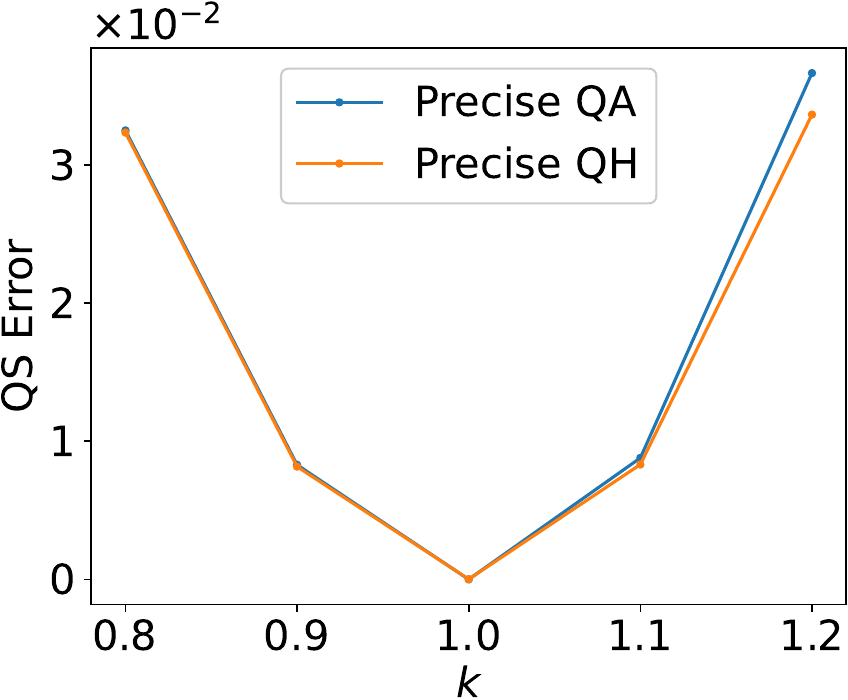}\hfill
    \includegraphics[width=0.51\linewidth]{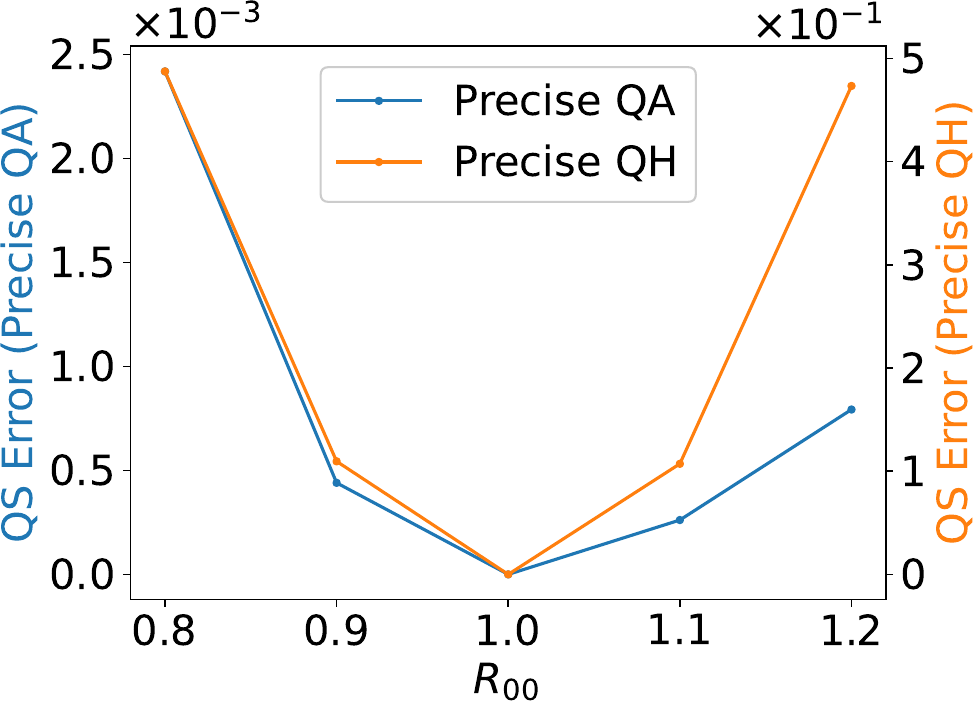}
    \caption{The quasisymmetry error, as defined in \eqref{eq:qse}, of the modified vacuum configurations based on the precise QA and QH configurations of \cite{LandremanPaul} a function of $k$ for the $\iota$ scan and of $R_{00}$ for the aspect ratio scan.}
    \label{fig:qse_plot}
\end{figure}

To verify these expectations numerically, we plot the quasisymmetry errors in the modified vacuum configurations based on the precise QA and QH in Figure \ref{fig:qse_plot}. In this figure, the quasisymmetry error is defined as
\begin{equation}
    \text{QS Error} = \frac{1}{V}\int_V \left[(N - \iota)({\bf B}\times\nabla B)\cdot\nabla\psi - G{\bf B}\cdot\nabla B\right]^2\frac{dV}{B^6},\label{eq:qse}
\end{equation}
where $V$ is the plasma volume. This is essentially the same metric that is used in the SIMSOPT code \citep{simsopt}, except that we take an average over the entire volume instead of weighing different flux surfaces differently.

\bibliography{bibfile}

\end{document}